\documentclass{elsart}
\usepackage{epsfig}

\def\Pm{{\mathcal P}}
\def\Pb{{\mathbf P}}

\def\beq{\begin{equation}}
\def\eeq{\end{equation}}
\def\bea{\begin{eqnarray}}
\def\eea{\end{eqnarray}}

\def\mki{\langle k_i \rangle}

\begin{document}

\begin{frontmatter}
\title{Preferential growth: Solution and application to modeling stock market}
\author[Bp]{L. Kullmann}
\author[Finn]{J. Kert\'esz}
\address[Bp]{Department of Theoretical Physics, 
Technical University of Budapest, Budafoki ut 8, H-1111, Budapest, Hungary}
\address[Finn]{Laboratory of Computational Engineering, 
Helsinki University of Technology,P.O.Box 9400, FIN-02015 HUT, Finland}
\begin{abstract}
We consider a preferential growth model where particles are added one
by one to the system consisting of clusters of particles. A new
particle can either form a new cluster (with probability $q$) or join
an already existing cluster with a probability proportional to the
size thereof. We calculate exactly the probability $\Pm_i(k,t)$ that
the size of the $i$-th cluster at time $t$ is $k$. We applied our
model as a background for a microscopic economic model.
\end{abstract}
\end{frontmatter}
\section{Introduction}
Preferential growth describes systems consisting of groups of entities
where the probability of the attachment of a new entity to one of the
groups is an increasing function of the group's size.  Recently these
types of models were used to describe networks like the World Wide Web
(WWW)~\cite{BARAB}, Internet~\cite{FAL}, statistics of scientific
citation~\cite{REDNER} {\it etc} which seems to have the common
property of scale invariance. It turned out that this behavior is due
to two main facts namely that they are continuously evolving and that
the attachment probability is proportional to the group's size.  Many
features of the above mentioned systems were analyzed and also
analytic calculations have been presented for the most important
quantities~\cite{DOROG} but they are only valid in the asymptotic time
case. Here we present the full-time dependent solutions.

We applied this kind of model to describe herding in an economic
system. It was suggested~\cite{CONT} that large price movements on the
stock market can be explained by the fact that traders are not
individual participants but they prefer to form groups, within a group
each trader share the same strategy. Our assumption is that those
groups are created according to a preferential growth
model. Preferential growth occurs naturally in
economics~\cite{AOKI}. The power-law behavior of the resulting group
distribution can account for the fat-tail distribution of the return
which is one of the stylized fact that characterize the stock market.

In the paper we present the full-time dependent solution of the growth
model, and briefly describe the most important results of our economic
model.
\section{Growth model}
The system consists of individual groups of entities. The initial
condition is one group with one entity in it. At each time step one new
entity is added to the system. With probability $q$ it
creates a new group, with $p = 1-q$ it will belong to one of the
existing groups. The probability that it joins the $i$th group is
proportional to the $i$th group's size ($k_i/N$).
\begin{figure}[ht]
\centerline{
\epsfig{file=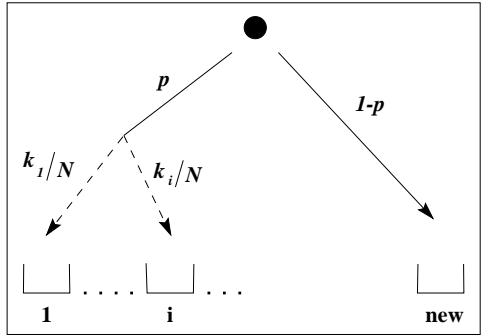,width=15pc}}
\caption{\footnotesize Illustration of the dynamic of our model.}
\label{model}
\end{figure}
The system's dynamic -- the time evolution of the $i$th group's size
 distribution, $\Pm_i(k,t)$ -- can be described by a master equation:
\begin{eqnarray}
\label{master}
\Pm_i(k,t)
&=& p{(k-1) \over t-1}\, \Pm_i(k-1,t-1)
+ p\left(1-{k \over t-1}\right)\, \Pm_i(k,t-1) + \nonumber\\
&&+\ (1-p)\, \Pm_i(k,t-1)
+ (1-p)\, \Pi_{i-1}(t-1)
\, \delta_{k,1}(1-\delta_{i,1}).
\end{eqnarray}
The quantity $\Pi_i(t)$ denotes the probability that at time $t$ the
system consists of $i$ group: $\Pi_i(t) = {t-1 \choose i-1}\
p^{t-1-(i-1)} (1-p)^{i-1}$.
Our goal is to determine the analytic full-time dependent form of the
following quantities: size distribution, $\Pm_i(k,t)$, and mean, $\mki
(t)$, of the individual groups and the average group size
distribution, $\Pb(k,t) = {\displaystyle {1 \over t} \sum_{i=1}^t}
\Pm_i(k,t)$. Here we do not go into the details of the deduction we
only present the results. (For more informations see\cite{KULL}).  For
the size distribution of the individual groups one gets:
\begin{eqnarray}
\label{Pm_i,k,t}
\Pm_i(k,t)\ &=&\
\sum_{l=1}^k\ (-1)^{l-1}\ {k-1 \choose l-1}\
{\Gamma(t-lp) \over \Gamma(t)\Gamma(1-lp)}
\times \nonumber \\
&&\qquad \times \left[\ \sum_{b=i}^{t}\ {\Gamma(b)\Gamma(1-lp) \over 
\Gamma(b-lp)}\ {b-2 \choose i-2} p^{b-i}\ (1-p)^{i-1} \right].
\end{eqnarray}
From $\Pm_i(k,t)$ we can easily calculate the analytic form of
$\Pb(k,t)$ and $\mki(t)$ by simple substitution in their definitions
and we get:
\begin{eqnarray}
\label{P_k,t}
\Pb(k,t) &=&
\sum_{l=1}^k\ (-1)^{l-1} {k-1 \choose l-1} \left[ {1-p \over 1+lp}
+ {p+lp \over 1+lp}\
{\Gamma(t-lp) \over \Gamma(t+1)\Gamma(1-lp)}
\right] \\
&& \nonumber \\
&& \nonumber \\
\mki (t) &=&
\sum_{l=1}^{t-i+1} (-1)^{l-1}\ l\ {t-i+2 \choose l+1}
{\Gamma(t-lp) \over \Gamma(t)\Gamma(1-lp)} \times \nonumber \\
&& \qquad \times 
\left[\ \sum_{b=i}^t {\Gamma(b)\Gamma(1-lp) \over \Gamma(b-lp)}\
{b-2 \choose i-2}\ p^{b-i}\ (1-p)^{i-1} \right] .
\end{eqnarray}

In the case when the system is
large ($t \to \infty$) it is worth to analyze the asymptotic
case of the above quantities in order to gain a much simpler form.

{\it i) Individual group size distribution}

In Eq. (\ref{Pm_i,k,t}) there are two time dependent terms. One is
$\displaystyle {\Gamma(t-lp) \over \Gamma(t)\Gamma(1-lp)}$, the other is
the sum in the bracket. For large $t$ values the second term will be
proportional to a hypergeometric function, $_2F_1(i,i-1;i-lp;p)$, and
will be time independent. The first term will converge to $t^{-lp}$ which
is a fast decaying function of $l$ so for $t \gg k$ one can
assume that only the first term of the sum in Eq. (\ref{Pm_i,k,t})
gives non-negligible component:
\begin{equation}
\label{Pm_i,k,inf}
\lim_{t \to \infty} \Pm_i(k,t) =
t^{-p}\ (1-p)^{i-1}
{\Gamma(i) \over \Gamma(i-p)}\ {_2}F_1 (i,i-1;i-p;p)
 + {\mathcal O}(t^{-2p})
\end{equation}
For large $i$ values the above form simplifies further and we finally get:
\begin{equation}
\label{Pm_inf,k,inf}
\lim_{t,i \to \infty} \Pm_i(k,t) = \left( {i \over t} \right)^p
\end{equation}
%

{\it ii) Mean of individual group size}

In the analysis of the mean value we assume that for $k \ll t$ the
distribution can be described by Eq. (\ref{Pm_i,k,inf}) for larger
$k$ values it has a fast decay so: $\displaystyle \mki (t) \approx
\sum_{k=1}^{k^*} k\ \Pm_i(1,t \to \infty)$. (The value $k^*$ can be
defined {\it e.g.} as the inflection point of $\Pm_i(k,t)$).
Here again analyzing the large $i$ case we get:

{\it iii) Distribution of average group size}

In the determination of the asymptotic case of the average group size
distribution, $\Pb(k)$, we can apply two ways. One is to simply
analyze the limit of Eq. (\ref{P_k,t}) or we can also start directly
from the master equation, Eq. (\ref{master}), by summing it up for $i
= 1 \dots t$. In both cases we get that in the long time limit the
distribution will be time independent and will have a power-law decay:
\begin{equation}
\label{P_k,inf}
\Pb(k) = {\Gamma(k)\Gamma \left(2+{1 \over p} \right) 
\over \Gamma \left(k+1+{1\over p} \right) }\,
{1-p \over 1+p}\quad \mathop{\sim}_{k \to \infty} k^{-1-1/p}.
\end{equation}
Here we would like to mention that however our growth model is not a
network model for particular values of the parameter ($p=0.5$) it can be
considered as a mean field analogy of the Barab\'asi's network
model. The resulting exponent of the distribution, ($-3$),
agrees with their result~\cite{BARAB}.
\section{Economic model}
The main properties of stock markets price movements are characterized
by the so called stylized facts, the distribution of the logarithmic
return fat-tailed, autocorrelation of the return is short range while
that of the absolute value of return decays with a power-law. We try
to reproduce these properties by a simple model.

Power-law decay of the distribution of the logarithmic return can be
explained by the herding behavior on the market \cite{CONT}.  Price
changes are assumed to be proportional to the excess demand -- the
difference between the amount of buying and selling orders. If traders
form groups so that within each group everyone shares the same
strategy than the logarithmic return can be expressed as:
$\displaystyle x \sim \sum_{i=1}^n s_i \phi_i$, where $\phi_i$ is the
strategy (to buy, to sell or not to trade), $s_i$ is the size of the
$i$th group.
If the number of groups that are active (trading) at a given time step is
small the distribution of the logarithmic return, $P(x)$ will be
similar to that of the groups' size distribution, $P(s)$. In our model
groups are created according to the above mentioned growth model. Due
to Eq. (\ref{P_k,inf}) $P(s)$ will have a power-law decay which
indicates that for law activity the distribution of logarithmic return
is also a power-law. Similar results were get by \cite{STAUFFER} but
they considered trading groups as two dimensional percolation
clusters.

\begin{figure}[ht]
\centerline{
\epsfig{file=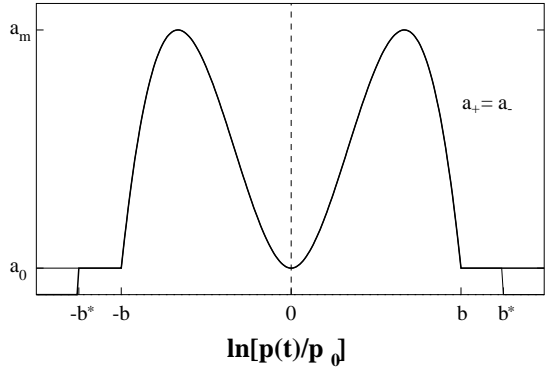,width=17pc}}
\caption{\footnotesize Price dependence of the activities. $a_+$ is
the fraction of the buying groups, $a_-$ is the fraction of selling
groups at a given time step. For $\ln[p(t)/p_0] < |b^*|$ the
activities $a_+$ and $a_-$ are equal. If $\ln[p(t)/p_0] > b^*$ then
$a_+ = 0,\ a_- = a_0$ while for $\ln[p(t)/p_0] < b^* \Rightarrow a_- =
0,\ a_+ = a_0$. $a_0$ is the reciprocal of the number of groups.}
\label{activity}
\end{figure}

Power-law correlation of the absolute value of price is due to the
long range correlation of the trading volume (number of trades in a
given time interval)~\cite{PLEROU}, which is in our model the sum of
the active groups' sizes, $S_a = \sum_{i=1}^{n_a} s_i$. We defined the
activity (fraction of groups that are trading at a given time step,
$n_a/n$) as the function of the deviation of the actual price from
some fundamental price, $p_0$; see Fig. \ref{activity}. (We took the
fundamental price for constant over time.) We ensure with this
definition that the activities of the proceeding time steps are
correlated.  For a given activity value the corresponding trading
volumes can be different but in average larger activity implies larger
trading volume which indicates that the sum of the trading groups'
sizes will also be correlated. The form of the function was chosen
intuitively, it is based on the hypothesis that small deviation from
the fundamental price inspire traders -- their activity rises -- while
for large deviations traders become careful so their activity falls
back.

\begin{figure}
\begin{minipage}[b][12pc][t]{16pc}
\epsfig{file=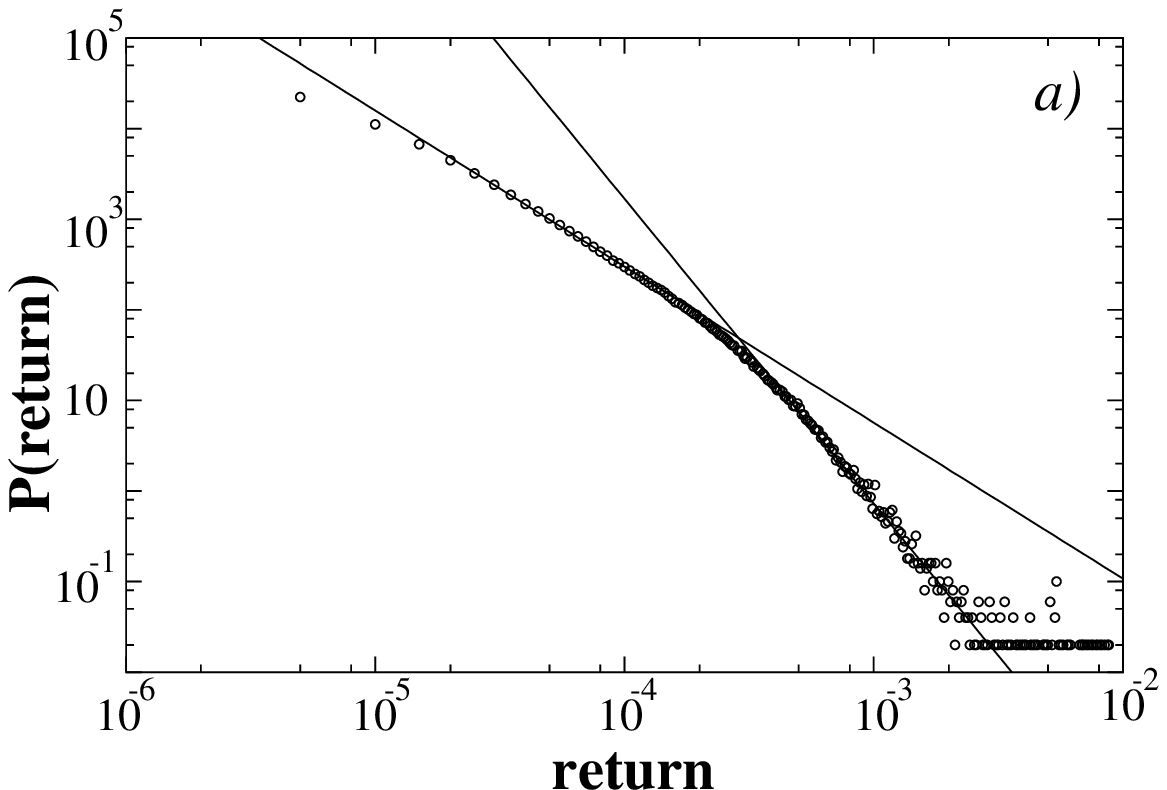,height=12pc,width=16pc}
\end{minipage}
\begin{minipage}[b][12pc][t]{16.5pc}
\epsfig{file=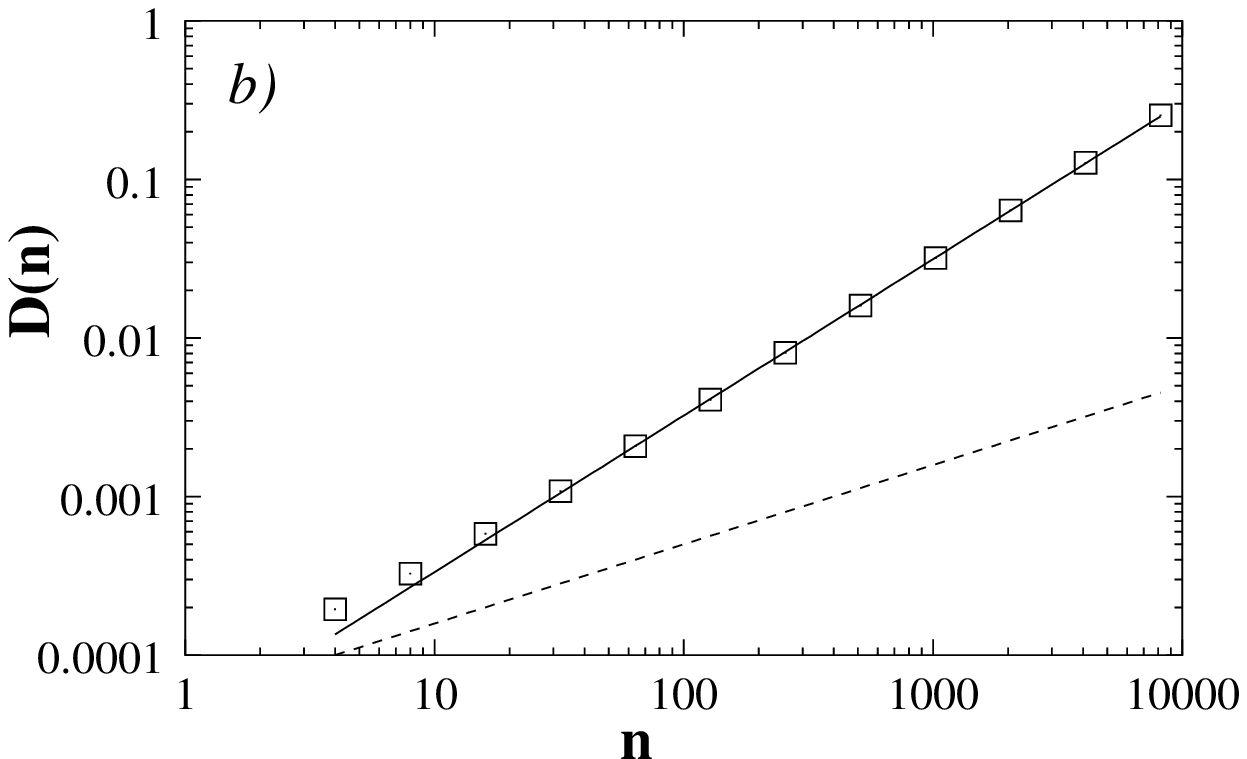,height=12pc,width=16.5pc}
\end{minipage}
\caption{\footnotesize {\it a}) Distribution of the logarithmic
return. It has on the log-log plot two approximately linear
regimes. The first has an exponent $\simeq 1.72$ the second $\simeq
3.36$ {\it b}) Square root of the variance, $D(n)$. The
dashed line corresponds to the uncorrelated, $\delta = 0.5$,
time series. The full line is a fit with exponent $\delta=0.95$.}
\label{result}
\end{figure}

The results of our model can be seen in Fig. \ref{result}. It is
surprising that for the distribution of the logarithmic return we not
only got a simple power-law but a distribution with two different
exponents (Fig. \ref{result}/a) which is similar to what was measured
on the real market\cite{GOPI}. This is, however, only a qualitative
agreement, the value of the exponents differ from the measured
ones~\cite{GOPI}. 
To analyze the autocorrelation of the absolute value of return we
study the variance of the sum of these variables (normalized to unit
variance):
\begin{eqnarray}
D^2(\sum_{i=1}^n x_i) \equiv D^2(n) = n + 2\sum_{m=1}^{n-1} (n-m)\
C(m) \sim n^{2\delta}, \nonumber
\end{eqnarray}
where $C(m) \sim m^{-\kappa}$ is the autocorrelation function. For
$\kappa \ge 1$ we have $\delta=0.5$, for variables with long-range
correlation ($\kappa <1$) $0.5 < \delta < 1$, and the correlation
exponent, $\kappa$, can be calculated through $\kappa = 2-2\delta$;
see Fig. \ref{result}/b.  It is clearly shown that the time series of
the absolute value of the logarithmic return is correlated but the
exponent is much smaller ($\simeq 0.05$) than the reported one
($\simeq 0.3$)~\cite{GOPI}.

The advantage of our model is its simplicity and that the results have
a qualitative agreement with the stylized facts. However, there are
tunable parameters in it and the results -- the values of the
exponents of the distribution and of the autocorrelation -- depend on
them. Further study is needed to clarify the role of these parameters.

{\bf Acknowledgment}: These research was supported by OTKA T029985.

\end{document}